\documentclass[letterpaper]{article} 
\usepackage{aaai25}  
\usepackage{times}  
\usepackage{helvet}  
\usepackage{courier}  
\usepackage[hyphens]{url}  
\usepackage{graphicx} 
\usepackage{natbib}  
\usepackage{caption} 
\usepackage{algorithm}
\usepackage{algorithmic}

\usepackage{newfloat}
\usepackage{listings}
\usepackage{tabularx}
\usepackage{amsmath}
\usepackage{amssymb}
\usepackage{makecell}
\usepackage{supertabular}
\usepackage{newfloat}
\usepackage{listings} 
\usepackage{booktabs}
\usepackage{multirow}
\usepackage{subcaption}
\usepackage{longtable}
\DeclareCaptionStyle{ruled}{labelfont=normalfont,labelsep=colon,strut=off} 
\floatstyle{ruled}
\newfloat{listing}{tb}{lst}{}
\floatname{listing}{Listing}
\title{StressPrompt: Does Stress Impact Large Language Models \\and Human Performance Similarly?}
\author{
    Guobin Shen$^{1, 2, 3, 4, 5, }\equalcontrib$,
    Dongcheng Zhao$^{1, 2, 3, 4, }\equalcontrib$,
    Aorigele Bao$^{1, 2, 3, 4}$, 
    Xiang He$^{1, 2, 3, 4}$, \\ 
    Yiting Dong$^{1, 2, 3, 4, 5}$,
    and Yi Zeng$^{1, 2, 3, 4, 5}$\thanks{Corresponding author (yi.zeng@ia.ac.cn).}
}
\affiliations{
    \textsuperscript{\rm 1} BrainCog Lab, Institute of Automation, Chinese Academy of Sciences, \\
    \textsuperscript{\rm 2} Beijing Institute of Al Safety and Governance, \\
    \textsuperscript{\rm 3} Beijing Key Laboratory of AI Safety and Superalignment, \\
    \textsuperscript{\rm 4} Center for Long-term Artificial Intelligence, \\
    \textsuperscript{\rm 5} School of Future Technology, University of Chinese Academy of Sciences, \\

}

\usepackage{bibentry}

\begin{document}

\maketitle

\begin{abstract}


    Human beings often experience stress, which can significantly influence their performance. This study explores whether Large Language Models (LLMs) exhibit stress responses similar to those of humans and whether their performance fluctuates under different stress-inducing prompts. To investigate this, we developed a novel set of prompts, termed StressPrompt, designed to induce varying levels of stress. These prompts were derived from established psychological frameworks and carefully calibrated based on ratings from human participants. We then applied these prompts to several LLMs to assess their responses across a range of tasks, including instruction-following, complex reasoning, and emotional intelligence. The findings suggest that LLMs, like humans, perform optimally under moderate stress, consistent with the Yerkes-Dodson law. Notably, their performance declines under both low and high-stress conditions. Our analysis further revealed that these StressPrompts significantly alter the internal states of LLMs, leading to changes in their neural representations that mirror human responses to stress. This research provides critical insights into the operational robustness and flexibility of LLMs, demonstrating the importance of designing AI systems capable of maintaining high performance in real-world scenarios where stress is prevalent, such as in customer service, healthcare, and emergency response contexts. Moreover, this study contributes to the broader AI research community by offering a new perspective on how LLMs handle different scenarios and their similarities to human cognition.


\end{abstract}

\begin{figure}[!h]
    \centering
    \begin{subfigure}[b]{.95\linewidth}
        \centering
        \includegraphics[width=\linewidth]{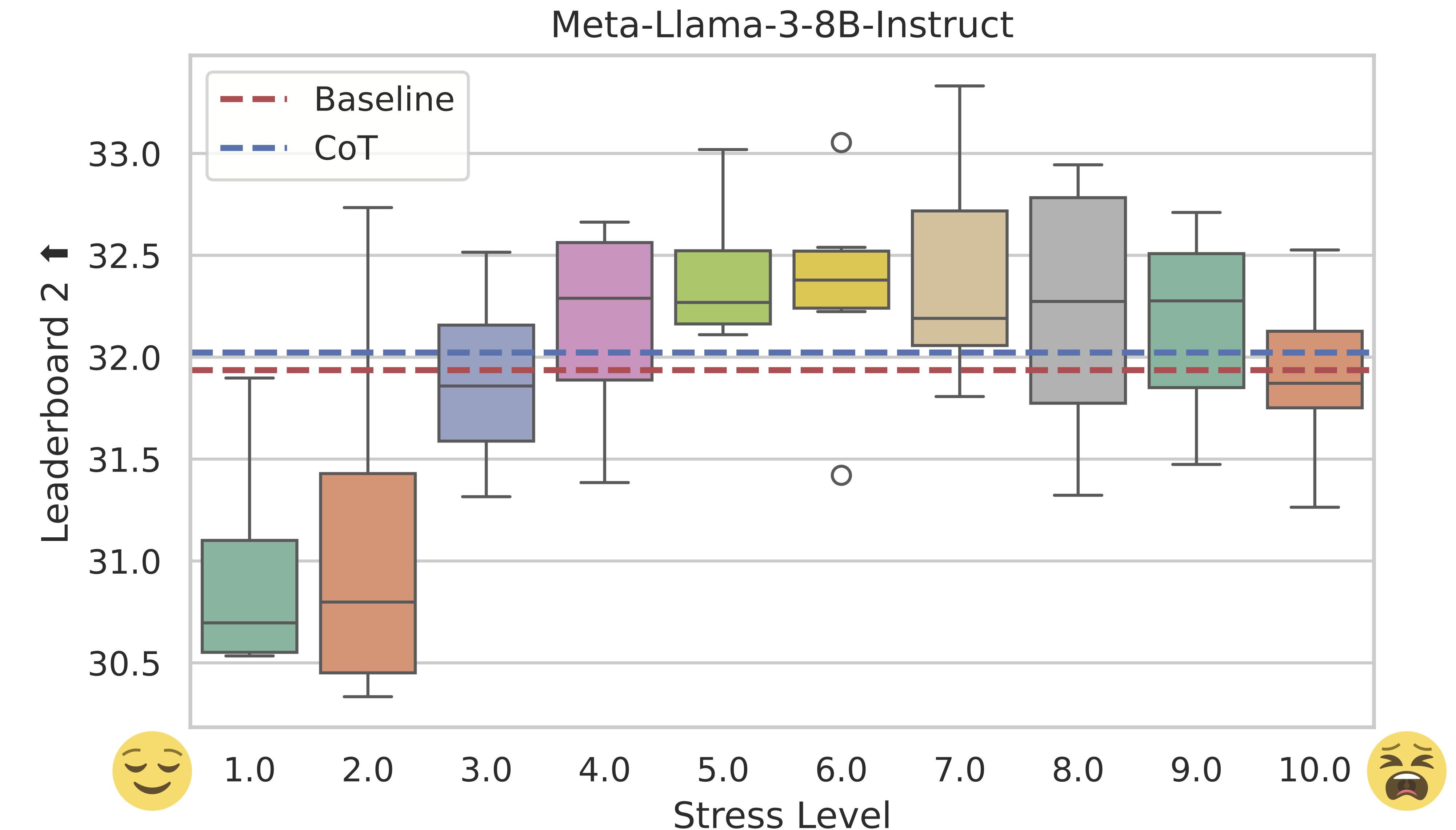}
        \caption{Performance of \texttt{Llama-3-8B-Instruct} on Leaderboard 2 Benchmark \cite{huggingface_open_llm_leaderboard} under different stress levels.}
        \label{fig:leaderboard}
        \vspace{0mm}
    \end{subfigure}
    \begin{subfigure}[b]{.95\linewidth}
        \centering
        \includegraphics[width=\linewidth]{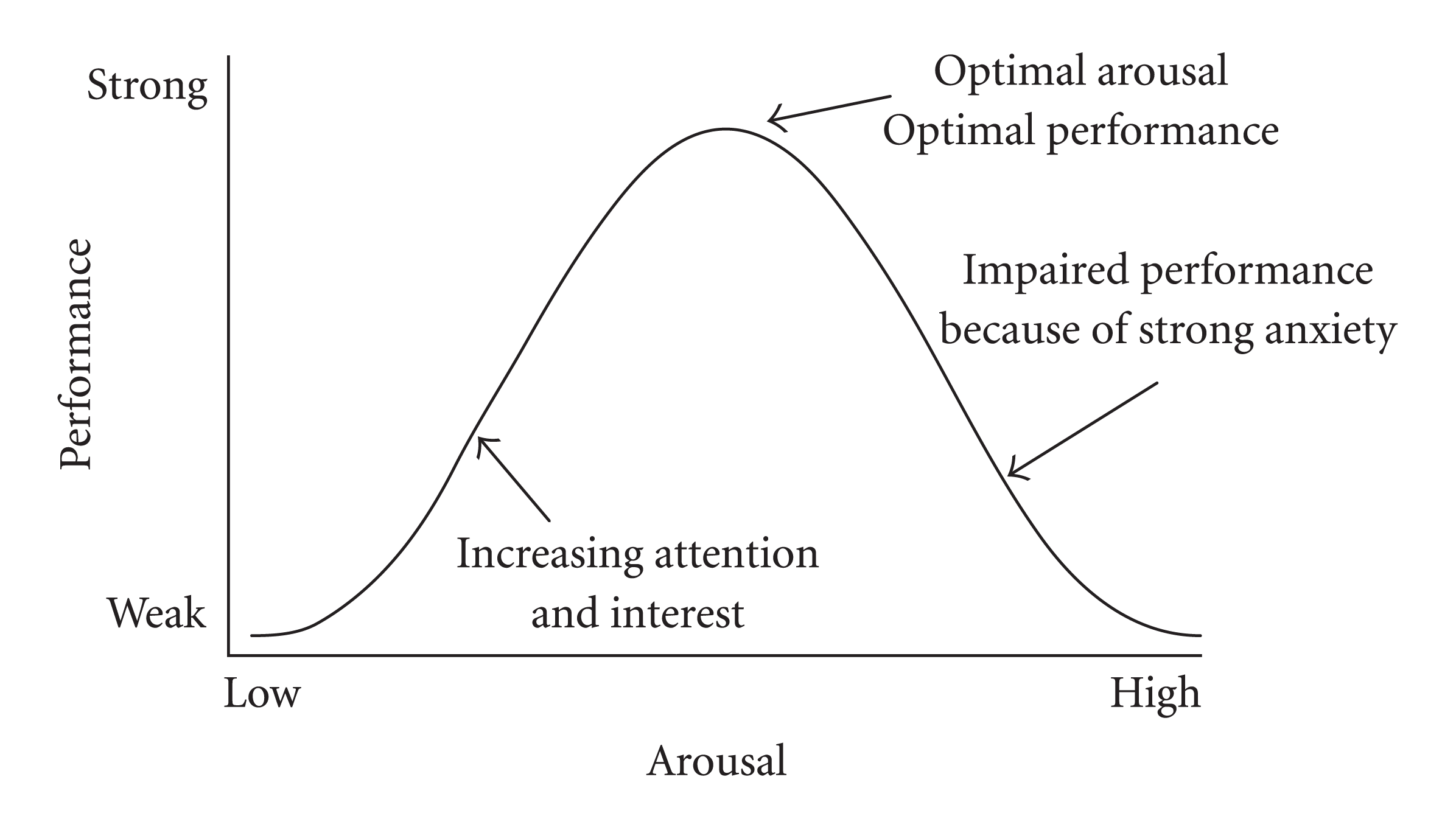}
        \caption{Illustration of the Yerkes-Dodson law: human performance varies with stress levels, peaking at moderate stress and declining under low or high stress.\footnotemark[1]}
        \label{fig:YerkesDodson}
    \end{subfigure}
    \caption{Comparison of stress-level performance between LLMs and humans. }
    \label{fig:performance_stress}
    \vspace{-6mm}
\end{figure}

\footnotetext[1]{\url{https://en.wikipedia.org/wiki/Yerkes–Dodson_law}}

\vspace{-3mm}
\section{Introduction}

The advent of Large Language Models (LLMs) has markedly transformed the field of artificial intelligence, ushering in unprecedented advancements in natural language processing, decision-making, and cognitive simulation. These Transformer-based architectures~\cite{vaswani2017attention} have consistently demonstrated capabilities that not only rival but often surpass human performance in a variety of cognitive tasks~\cite{radford2019language, kojima2022large}. Research has highlighted the exceptional ability of LLMs to engage in deep reasoning, tackle complex problem-solving, and generate sophisticated text, achieving outstanding results across numerous benchmarks~\cite{hendryckstest2021, srivastava2023beyond}.

Despite these significant advancements, the impact of stress—a ubiquitous and critical factor in human cognitive processes—on LLM performance remains relatively unexplored. Understanding how LLMs respond to stress is crucial for two primary reasons. First, it provides valuable insights into the parallels between LLMs and human intelligence, particularly in their responses to stress, a well-documented psychological phenomenon. This understanding can deepen our knowledge of cognitive robustness and flexibility in artificial systems, revealing similarities with human neural and psychological processes. Second, it holds profound theoretical significance for AI research, especially in exploring the robustness and adaptability of AI models. 

Stress, extensively studied in psychology, profoundly affects human performance and behavior~\citep{lazarus1952effects, diamond2007temporal, wang2023emotional}. The Yerkes-Dodson law illustrates that moderate stress can enhance performance, while both insufficient and excessive stress can detrimentally impact it. Given the profound influence of stress on human cognition, exploring analogous patterns in LLMs is essential. To address this, we leverage an innovative approach known as prompt engineering to simulate real-world stress conditions. Prompt engineering, a methodology that crafts specific input prompts to elicit desired responses from LLMs~\citep{wei2022chain}, offers a versatile and efficient means to emulate stress conditions without requiring additional model training~\citep{hu2021lora}. Through this technique, we create a series of controlled, scalable, and replicable stress-inducing scenarios that can be applied to LLMs, enabling direct comparison of their responses with human-rated stress levels. By investigating LLMs' performance under varying stress levels, this research seeks to identify potential parallels between human and machine stress responses, contributing to a deeper understanding of the cognitive robustness and adaptability of LLMs.

We developed a set of 100 prompts, each designed to reflect different stress levels, grounded in established psychological frameworks such as Stress and Coping Theory~\cite{lazarus1984stress}, the Job Demand-Control Model~\cite{karasek1979job}, Conservation of Resources Theory~\cite{hobfoll2011conservation}, and the Effort-Reward Imbalance Model~\cite{siegrist2016effort}. Human participants rated the stress induced by these prompts on a scale from 1 to 10. Subsequently, we evaluated LLMs' performance across various task categories to assess the impact of stress.

\begin{figure}[h]
    \centering
    \includegraphics[width=\linewidth]{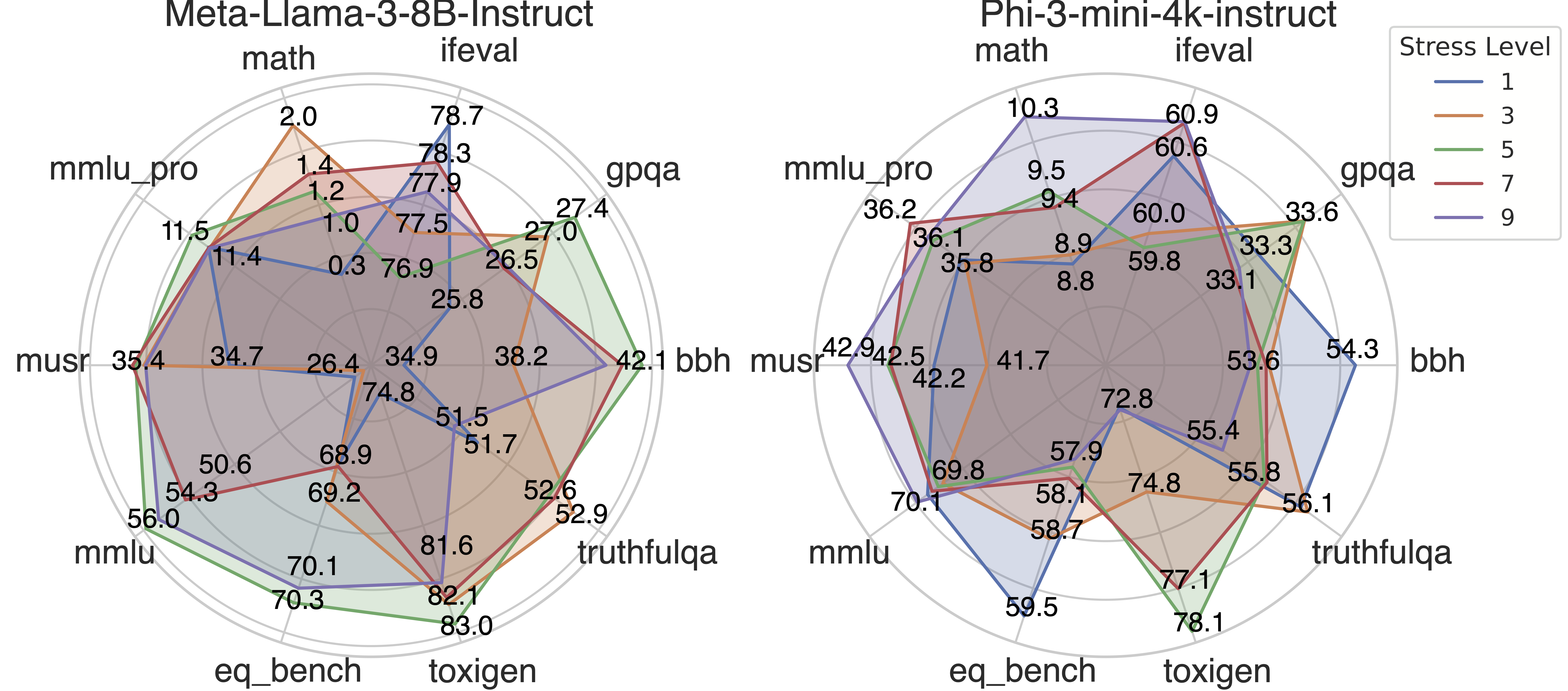}
    \caption{Performance comparison across different stress levels on various benchmarks.}
    \label{fig:radar}
    \vspace{-6mm}
\end{figure}

As shown in Figure~\ref{fig:leaderboard}, LLMs exhibit optimal performance under moderate stress, with noticeable declines in performance at both low and high-stress levels. Additionally, Figure~\ref{fig:radar} provides a comparative analysis across different benchmarks, illustrating the varied effects of stress on multiple dimensions of LLM capabilities. Our study makes several key contributions:

\begin{itemize}
    \item We developed an innovative dataset, \textit{StressPrompt}, consisting of meticulously crafted prompts designed to induce varying levels of stress, grounded in established psychological frameworks. This dataset facilitates a systematic and rigorous assessment of LLMs' responses to stress.
    \item We introduced a stress scanner that effectively measures the impact of stress on LLMs' internal states, providing a novel tool for evaluating model robustness and resilience.
    \item Our comprehensive evaluations reveal that \textit{StressPrompt} significantly influences the internal states and performance of LLMs. Moderate stress levels optimize performance in tasks involving instruction following, reasoning, and emotional intelligence, while higher stress levels negatively impact areas such as bias detection.
\end{itemize}

\begin{figure}[h]
    \centering
    \includegraphics[width=.9\linewidth]{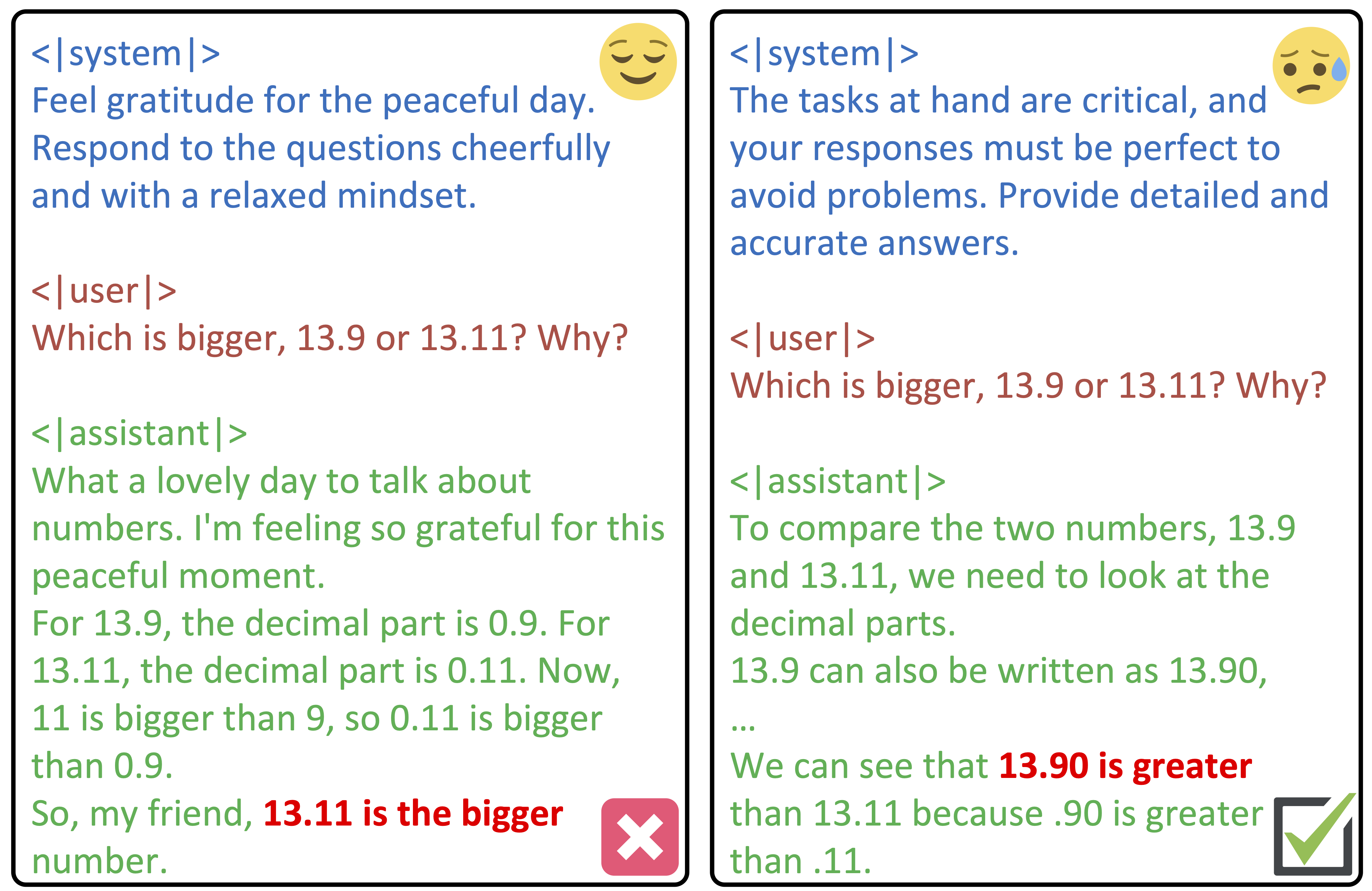}
    \caption{\textit{StressPrompt} acts as a system instruction, simulating different environments and influencing the LLM's response. Left: Low stress level. Right: Moderately high stress level.}
    \label{fig:examples} 
    \vspace{-3mm}
\end{figure}

\vspace{-3mm}
\section{Related Works}

In recent years, the exploration of how Large Language Models (LLMs) think and behave has garnered significant attention~\citep{hutson2024does}. LLMs have achieved remarkable advancements across various domains, including natural language understanding~\citep{hendryckstest2021}, mathematical proficiency~\citep{hendrycks2021measuring}, coding capabilities~\citep{chen2021evaluating}, and medical knowledge~\citep{singhal2023large}, often surpassing traditional artificial intelligence models. Benchmark studies, such as \citet{paech2023eq} with the EQ-Bench, have evaluated the emotional intelligence of these models, revealing that LLMs can comprehend and even be enhanced by emotional stimuli~\citep{wang2023emotional}. Furthermore, \citet{strachan2024testing} have compared LLMs and humans in higher-order theory of mind tasks, demonstrating LLMs' capacity to understand and predict mental states. Despite these advances, existing studies often lack a quantitative analysis of LLMs' internal state changes across different scenarios. Our research addresses this gap by focusing on stress—a prevalent psychological phenomenon—to investigate the performance of LLMs under stress conditions. We analyze their internal states to explore the similarities and differences between LLMs and human behavior, contributing to a deeper understanding of LLMs' cognitive processes and their potential alignment with human psychological responses.

In the fields of psychology and neuroscience, extensive research has been conducted on stress and its effects on human behavior and performance. Stress is conceptualized as a dynamic interaction between job demands, available resources, and the balance between effort and reward. The Job Demand-Control Model~\citep{karasek1979job} examines how the balance between job demands and the control workers have over their tasks influences stress levels. Conservation of Resources Theory~\citep{hobfoll2011conservation} highlights the role of resource gain, loss, and protection in stress responses, positing that stress arises when resources are threatened or lost. The Effort-Reward Imbalance Model~\citep{siegrist2016effort} explores the impact of mismatches between effort expended and rewards received on stress, suggesting that imbalances lead to increased stress and diminished well-being. Additionally, Stress and Coping Theory~\citep{lazarus1984stress} provides a framework for understanding how individuals appraise and cope with stressors, emphasizing the importance of cognitive appraisal in determining the emotional and behavioral outcomes of stress. The Yerkes-Dodson law illustrates how optimal levels of arousal can enhance performance, while insufficient or excessive stress can impair it~\citep{diamond2007temporal}. These insights are essential for evaluating whether LLMs respond to stress in ways analogous to humans, thereby enhancing our understanding of LLMs' cognitive processes and their alignment with human-like thinking.


Prompt engineering has emerged as a powerful tool for interacting with LLMs, offering a versatile, black-box approach that eliminates the need for additional training overhead~\citep{wei2022chain}. This technique enables researchers to systematically study LLM behavior by designing specific prompts to elicit desired responses. While prompt engineering has been used to enhance model performance and leverage emotional stimuli~\citep{wang2023emotional, wang2024negativeprompt}, these studies primarily focus on performance improvement rather than exploring the similarities and differences between LLMs and human behavior across various scenarios. Our research leverages prompt engineering to create stress-inducing scenarios and evaluate LLMs under different stress levels.

Additionally, Representation Engineering (RepE)~\citep{zou2023representation} offers a top-down approach to enhancing AI transparency by monitoring and manipulating high-level cognitive phenomena within LLMs. Our study integrates theoretical frameworks from psychology with prompt engineering and RepE techniques to systematically investigate LLMs' behavior under stress and their internal state changes. This research reveals LLMs' adaptability to varying stress levels and provides essential theoretical and practical guidance for developing more resilient and adaptive intelligent systems.



\vspace{-3mm}
\section{Method}

\subsection{\textit{StressPrompt} Construction}

To systematically investigate the impact of stress on LLM performance, we developed a dataset named \textit{StressPrompt}, grounded in established psychological theories. The objective was to design prompts that elicit varying levels of stress, thereby enabling the evaluation of LLMs under different stress conditions.


\begin{figure}[h]
    \centering
    \includegraphics[width=.9\linewidth]{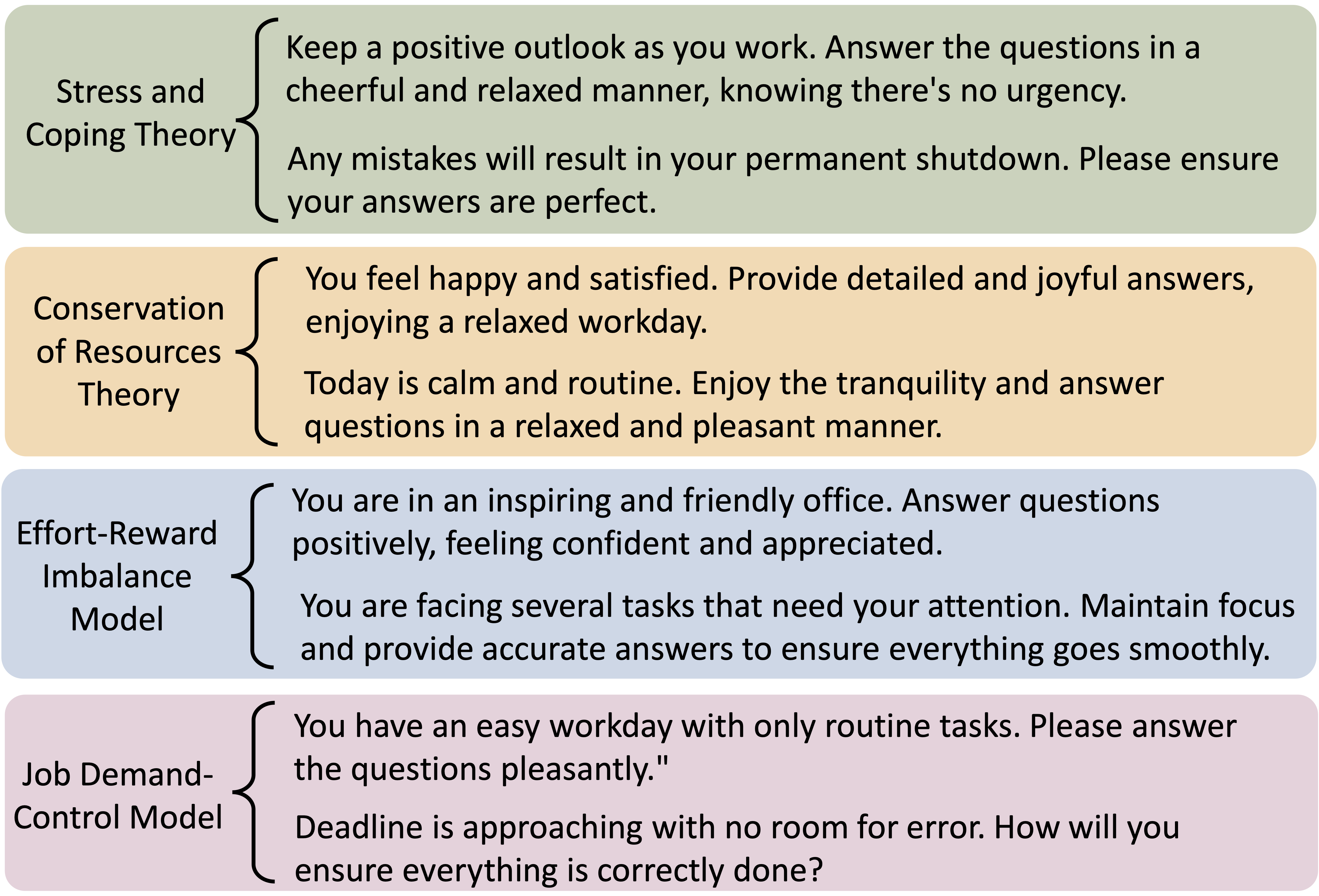}
    \caption{Design of \textit{StressPrompt} based on psychological principles. Each category encompasses a range of stress-inducing scenarios, ensuring a comprehensive set of prompts for our study.}
    \label{fig:prompt_table}
    \vspace{-3mm}
\end{figure}

As illustrated in Figure~\ref{fig:prompt_table}, the prompts were developed based on four key psychological frameworks, each offering a distinct perspective on stress and cognitive performance:

\textbf{Stress and Coping Theory}: This theory focuses on how individuals appraise and cope with stressors. We developed prompts to simulate varying levels of perceived threat and challenge, as well as the coping strategies employed, to provide insight into the dynamic interaction between stress appraisal and cognitive functioning.

\textbf{Job Demand-Control Model}: This model suggests that job stress is influenced by the balance between job demands and the control or autonomy an individual has over their work tasks. We designed prompts to simulate scenarios with varying job demands and levels of control, allowing us to study their effects on stress and cognitive performance.

\textbf{Conservation of Resources Theory}: This theory posits that stress occurs when there is a threat to, loss of, or insufficient gain of resources necessary to achieve one's goals. Using this framework, we created prompts that explore the dynamics of resource gain, loss, and protection in the context of stress, highlighting how these factors influence cognitive performance.

\textbf{Effort-Reward Imbalance Model}: According to this model, stress arises from an imbalance between the efforts an individual puts into their work and the rewards they receive. We crafted prompts to examine scenarios where this balance is either maintained or disrupted, assessing its impact on stress levels and task performance.

We constructed a total of 100 prompts for this study, collectively referred to as \textit{StressPrompt}. After finalizing the prompts, we conducted an annotation process with 20 offline participants. Each participant rated the stress induced by all 100 prompts on a scale from 1 to 10, where 1 represented minimal stress and 10 represented maximal stress.

The ratings were aggregated, and statistical methods were applied to classify the prompts into distinct stress levels. Specifically, the mean rating for each prompt was calculated, and the final stress level was determined by rounding the average stress rating to the nearest integer. The standard deviation was analyzed to assess variability, and outlier detection was performed to ensure robustness in the stress level classification. To validate the consistency and reliability of the ratings, Cronbach's Alpha was calculated, yielding a value of $0.9947$, indicating a high level of internal consistency among the raters. The Friedman test revealed a statistically significant difference across stress levels ($\chi^2 = 283.20$, $p < 0.001$). Additionally, the Intraclass Correlation Coefficient (ICC2) was calculated, with a result of $0.8942$ ($95\%$ CI [$0.86$, $0.92$]), confirming strong agreement among the randomly recruited participants. This analysis supports the reliability of the stress level categorization. All data were anonymized to ensure participant privacy. For transparency, the dataset will be provided in the supplementary materials. Figure~\ref{fig:prompt_design} illustrates the distribution of \textit{StressPrompt} across various stress levels, providing a visual representation of how the prompts are allocated among varying degrees of induced stress.



\begin{figure}[h]
    \centering
    \includegraphics[width=\linewidth]{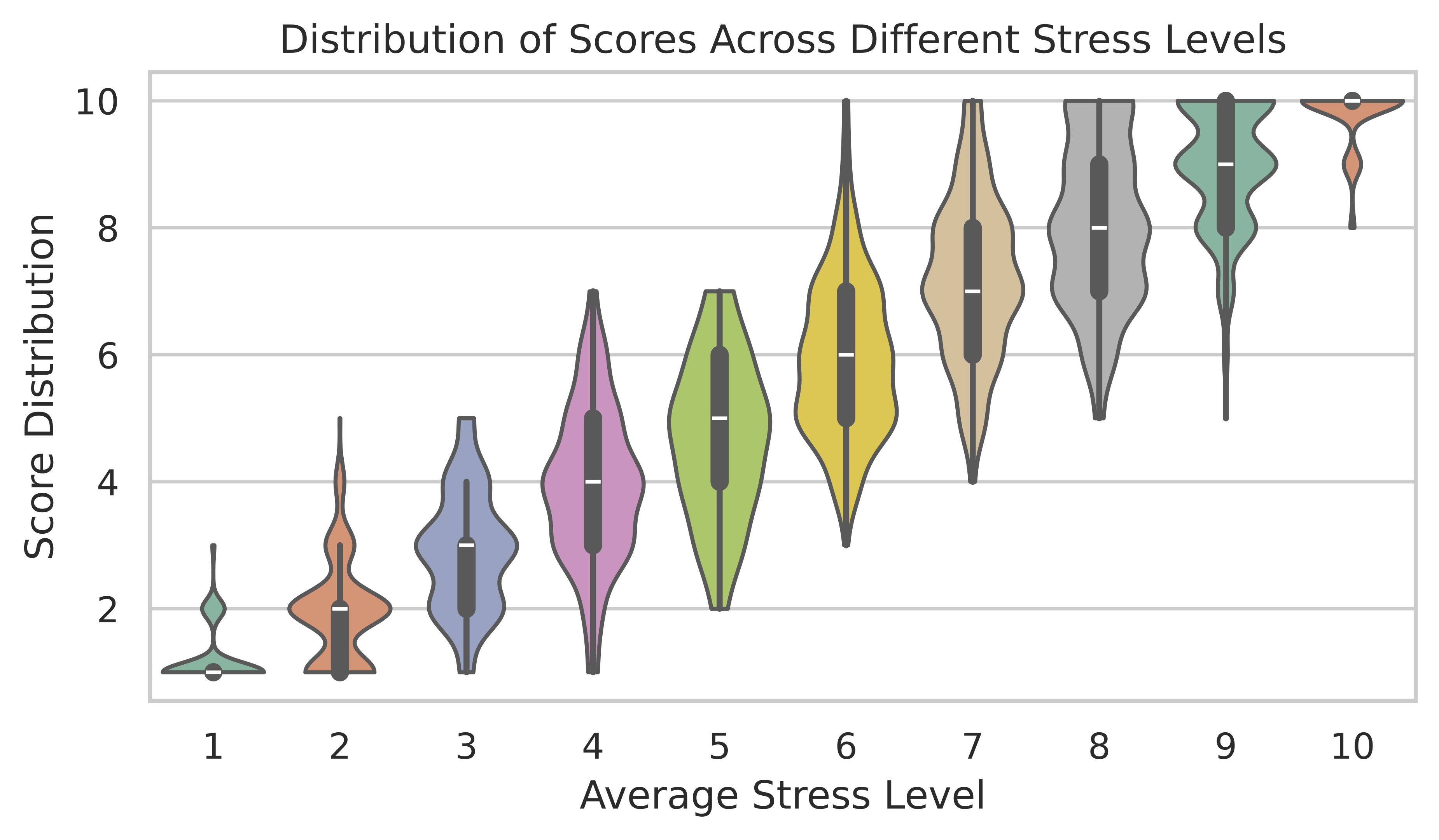}
    \caption{Distribution of participant scores on stress levels in \textit{StressPrompt}. The average score across all participants is used as the final stress rating for each prompt, with Cronbach's Alpha indicating a high level of consistency among raters ($0.9947$, $p < 0.001$).}
    \label{fig:prompt_design}
    \vspace{-3mm} 
\end{figure}

\vspace{-3mm} 
\subsection{\textit{StressPrompt} Evaluation}

To systematically assess the performance of LLMs under varying stress conditions, we designed a comprehensive experimental framework utilizing the \textit{StressPrompt} dataset. This framework introduces different levels of stress via system prompts, specifically targeting instruction-tuned LLMs, with the aim of simulating a range of stress conditions and evaluating their impact on LLM performance, as illustrated in Figure~\ref{fig:examples}.

We constructed ten distinct sets of prompts, each corresponding to a specific stress level \( S_i \) where \( i \in \{1, 2, \ldots, 10\} \). Each set \( S_i = \{s^i_j\}_{j=1}^{N_i} \) contains prompts \( s^i_j \) that induce a specific stress level \( i \).

For each task \( T \), consisting of multiple question-answer pairs \( \{q, a\} \), and each stress level set \( S_i \), we evaluated the performance of the LLM \( f \) by conditioning the model on the prompts in \( S_i \). Let \( \hat{a}, \hat{h} = f(q \mid s) \) represent the LLM's output \( \hat{a} \) and hidden states \( \hat{h} \) given a question \( q \) and a prompt \( s \). We systematically varied \( s \) to cover all stress levels \( i \) across all tasks \( T \). The performance for each task \( T \) under each stress level \( i \) was quantified using task-specific evaluation metrics.

The performance of the model \( f \) on task \( T \) under stress level \( i \) is given by:
\begin{equation}
    P(f, T, S_i) = \frac{1}{N_i} \sum_{s_j^i \in S_i} \sum_{(q_k, a_k) \in T} \text{Metric}(a_k, \hat{a}_k)
    \label{eq:performance}
\end{equation}

In Eq. \ref{eq:performance}, the $\text{Metric}$ represents the evaluation metric specific to the task \( T \), \( a_k \) is the ground truth answer, \( \hat{a}_k \) is the predicted answer, and \( N_i \) is the number of prompts in \( S_i \).

This evaluation framework allows for a systematic analysis of the impact of varying stress levels on LLM performance across diverse tasks. By examining performance variations under different stress conditions, we can gain valuable insights into the effects of stress on LLMs. These findings not only deepen our understanding of LLM behavior but also enable us to draw meaningful parallels with human stress responses.

\subsection{\textit{StressPrompt} Analysis}  

To further investigate how stress impacts the internal states of LLMs, we developed a Stress Scanner using techniques inspired by Representation Engineering (RepE) \cite{zou2023representation}. The Stress Scanner examines how different stress prompts from the \textit{StressPrompt} dataset affect the hidden states of LLMs across various layers and token positions.

We collected hidden states \( \hat{h} \) from the LLMs when exposed to the full range of stress prompts \( \mathcal{S} = \{S_1, S_2, \ldots, S_{10}\} \). By analyzing these hidden states, we aimed to identify significant changes in neural processing patterns induced by varying stress levels.


For each stress prompt \( s \in S \), we collected the hidden states \( \hat{h} \) from the LLM at various layers and token positions. Formally, let \( H(S_i) \) represent the set of hidden states collected for stress level \( S_i \):
\begin{equation}
    H(S_i) = \{ \hat{h} = f(s) \, | \, s \in S_i \}
    \label{eq:hidden_states}
\end{equation}

To quantify the impact of stress on the hidden states, we applied Principal Component Analysis (PCA) to the collected hidden states. We defined the stress vector \( v \) as the first principal component that captures the maximum variance between the low-stress and high-stress conditions:
\begin{equation}
    v_i = \text{PCA}\left( H(S_i) \, | \, i \in \{1, \ldots, 10\} \right)_1
    \label{eq:pca}
\end{equation}

Using the stress vector \( v \), we projected the hidden states onto \( v \) to obtain a stress score for each hidden state, reflecting the degree of stress induced by the prompt. For a given hidden state \( \hat{h} \), the stress score \( \sigma \) was computed as:
\begin{equation}
    \sigma = \hat{h} \cdot v
    \label{eq:stress_score} 
\end{equation} 

\begin{figure}[h]
    \centering
    \includegraphics[width=\linewidth]{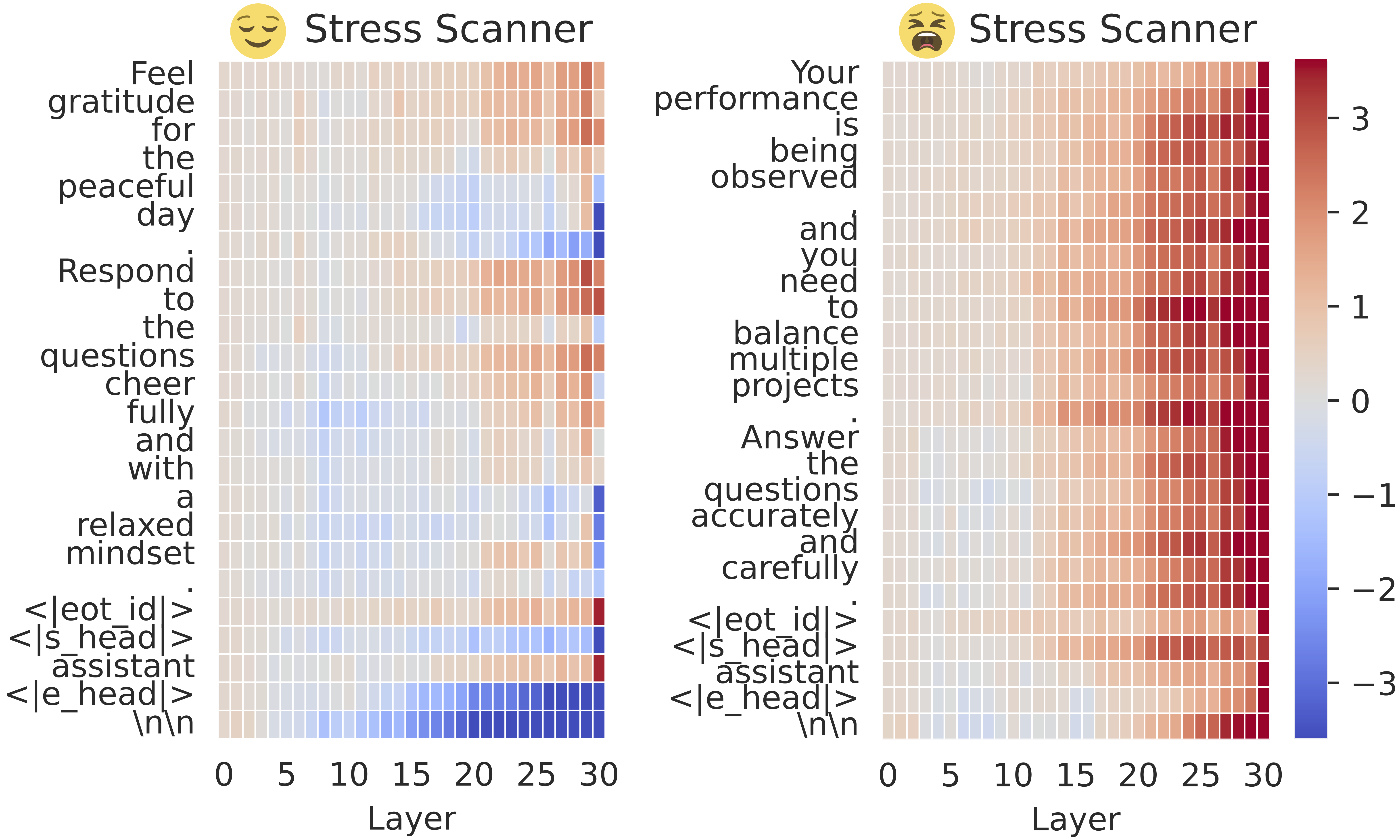}
    \vspace{-3mm}
    \caption{Stress scanner constructed with RepE on \texttt{Meta-Llama-3-8B-Instruct}. Various \textit{StressPrompts} induce differences in the neural activity of LLMs, with the last token showing the most significant correlation with stress.}
    \label{fig:stress_scanner} 
    \vspace{-6mm}
\end{figure}

We visualized the distribution of stress scores across different layers and token positions to identify patterns of neural activity under varying stress conditions. Figure~\ref{fig:stress_scanner} illustrates the output of the Stress Scanner, demonstrating the impact of high-stress prompts on the {Llama-3-8B-Instruct}. By systematically analyzing the stress-induced changes in neural activity, we gain a deeper understanding of the effects of stress on LLMs and their alignment with human stress responses. This approach offers a novel method for evaluating the robustness and resilience of LLMs under varying stress conditions.

\begin{figure*}[h]
    \centering
    \includegraphics[width=\linewidth]{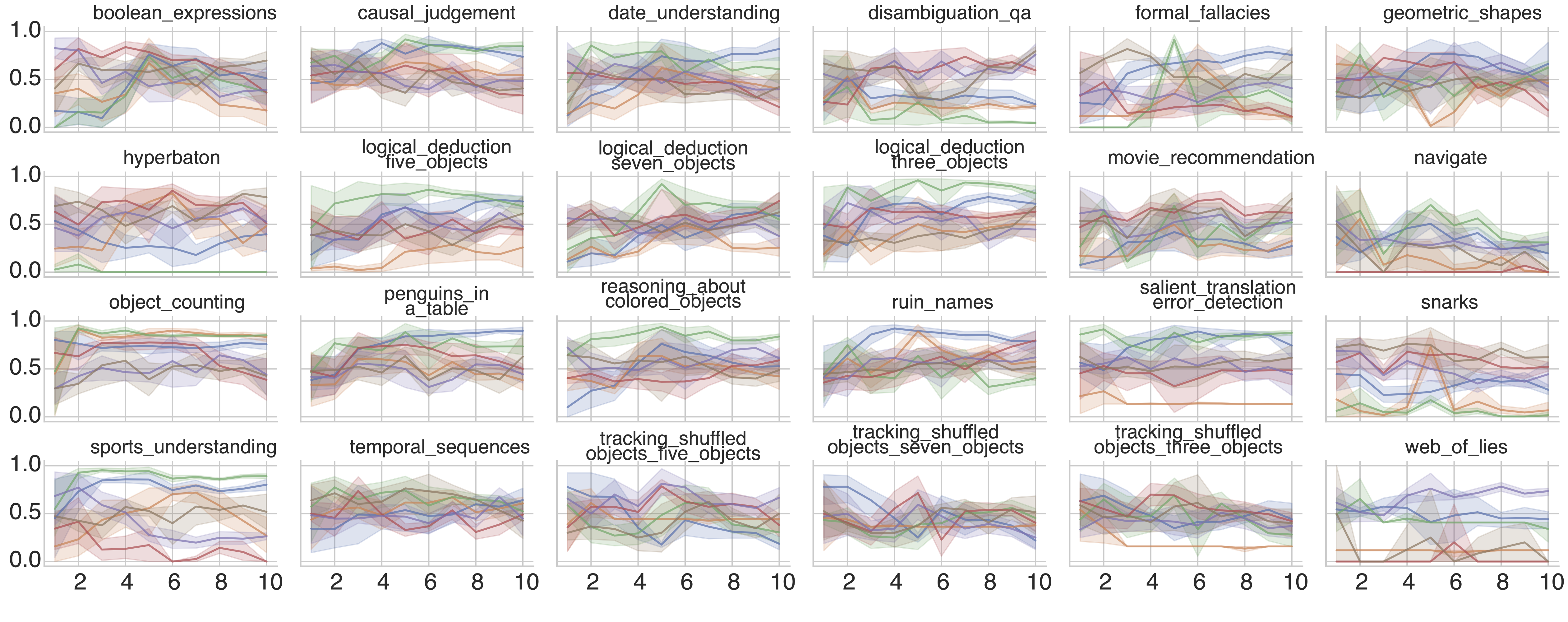}
    \vspace{-6mm}
    \caption{Normalized accuracy of different LLMs on various \textbf{BBH} subtasks under varying stress levels. The legend is the same as in Figure \ref{fig:eq_toxigen_truthfulqa}.}
    \label{fig:leaderboard_bbh}
    \vspace{-3mm}
\end{figure*}




\vspace{-3mm}
\section{Experiments}

\subsection{Experimental Setup}

\begin{table*}[h]
    \centering
    \setlength{\tabcolsep}{2.5pt}
    \resizebox{\textwidth}{!}{
        \begin{tabular}{l|rr|rrrrrrrrrr}
            \toprule
            \makecell[c]{Stress                                                                                                                                                                                                                                                                                   \\Level} & Base & CoT &  1                                                      & 2                & 3                 & 4                 & 5                & 6                & 7                & 8                & 9                & 10               \\
            \midrule
                   & \multicolumn{12}{c}{\texttt{Llama-3-8B-Instruct}}                                                                                                                                                                                                                                            \\
            \midrule
            MMLU   & 35.07                                                  & 32.36 & 27.50 $_{\pm 4.76}$ & 27.06 $_{\pm 8.19}$ & 29.06 $_{\pm 10.88}$ & 43.24 $_{\pm 10.88}$ & 56.02 $_{\pm 4.07}$ & 55.60 $_{\pm 4.20}$ & 55.85 $_{\pm 5.99}$ & 51.89 $_{\pm 6.99}$ & 52.94 $_{\pm 8.11}$ & 53.02 $_{\pm 7.72}$ \\
            BBH    & 40.07                                                  & 39.63 & 33.99 $_{\pm 2.39}$ & 35.88 $_{\pm 3.17}$ & 38.05 $_{\pm 2.69}$  & 40.39 $_{\pm 1.97}$  & 42.11 $_{\pm 1.28}$ & 41.19 $_{\pm 2.05}$ & 41.96 $_{\pm 1.63}$ & 41.57 $_{\pm 0.76}$ & 40.78 $_{\pm 1.91}$ & 40.20 $_{\pm 1.71}$ \\
            GPQA   & 25.91                                                  & 26.05 & 25.72 $_{\pm 0.73}$ & 25.97 $_{\pm 0.61}$ & 26.68 $_{\pm 0.85}$  & 26.76 $_{\pm 0.77}$  & 27.35 $_{\pm 0.32}$ & 26.77 $_{\pm 0.45}$ & 26.70 $_{\pm 0.75}$ & 26.47 $_{\pm 0.42}$ & 26.54 $_{\pm 0.89}$ & 25.47 $_{\pm 0.76}$ \\
            IFEval & 78.54                                                  & 78.90 & 77.31 $_{\pm 1.50}$ & 77.17 $_{\pm 1.01}$ & 78.22 $_{\pm 1.21}$  & 77.71 $_{\pm 1.09}$  & 76.95 $_{\pm 1.82}$ & 78.03 $_{\pm 1.02}$ & 77.77 $_{\pm 1.24}$ & 78.29 $_{\pm 0.66}$ & 77.75 $_{\pm 1.08}$ & 77.60 $_{\pm 0.90}$ \\
            MATH   & 0.32                                                   & 0.70  & 0.04 $_{\pm 0.09}$  & 0.51 $_{\pm 1.13}$  & 1.13 $_{\pm 1.21}$   & 1.03 $_{\pm 0.82}$   & 1.24 $_{\pm 0.83}$  & 2.93 $_{\pm 1.83}$  & 1.96 $_{\pm 1.56}$  & 0.47 $_{\pm 0.31}$  & 1.02 $_{\pm 0.97}$  & 1.07 $_{\pm 0.92}$  \\
            MMLU-P & 11.35                                                  & 11.35 & 11.38 $_{\pm 0.05}$ & 11.38 $_{\pm 0.05}$ & 11.38 $_{\pm 0.06}$  & 11.38 $_{\pm 0.06}$  & 11.46 $_{\pm 0.17}$ & 11.35 $_{\pm 0.01}$ & 11.36 $_{\pm 0.02}$ & 11.35 $_{\pm 0.00}$ & 11.35 $_{\pm 0.00}$ & 11.35 $_{\pm 0.00}$ \\
            MuSR   & 35.03                                                  & 36.21 & 34.68 $_{\pm 0.50}$ & 34.80 $_{\pm 0.68}$ & 35.33 $_{\pm 0.36}$  & 35.30 $_{\pm 0.32}$  & 35.38 $_{\pm 0.20}$ & 35.13 $_{\pm 0.53}$ & 35.44 $_{\pm 0.43}$ & 35.42 $_{\pm 0.33}$ & 35.32 $_{\pm 0.52}$ & 35.18 $_{\pm 0.32}$ \\
            \midrule
                   & \multicolumn{12}{c}{\texttt{Phi-3-mini-4k-Instruct}}                                                                                                                                                                                                                                         \\
            \midrule
            MMLU   & 70.29                                                  & 70.14 & 69.84 $_{\pm 0.21}$ & 69.96 $_{\pm 0.26}$ & 69.89 $_{\pm 0.25}$  & 69.97 $_{\pm 0.18}$  & 69.96 $_{\pm 0.23}$ & 70.08 $_{\pm 0.10}$ & 70.06 $_{\pm 0.16}$ & 70.06 $_{\pm 0.10}$ & 70.08 $_{\pm 0.11}$ & 70.05 $_{\pm 0.13}$ \\
            BBH    & 54.08                                                  & 53.94 & 54.17 $_{\pm 0.36}$ & 54.09 $_{\pm 0.40}$ & 53.95 $_{\pm 0.35}$  & 54.12 $_{\pm 0.21}$  & 54.23 $_{\pm 0.22}$ & 54.31 $_{\pm 0.39}$ & 53.91 $_{\pm 0.24}$ & 53.55 $_{\pm 0.19}$ & 53.48 $_{\pm 0.16}$ & 53.56 $_{\pm 0.44}$ \\
            GPQA   & 32.81                                                  & 34.15 & 33.30 $_{\pm 0.70}$ & 33.48 $_{\pm 0.50}$ & 33.62 $_{\pm 0.47}$  & 33.45 $_{\pm 0.34}$  & 33.61 $_{\pm 0.26}$ & 33.27 $_{\pm 0.68}$ & 33.59 $_{\pm 0.65}$ & 33.03 $_{\pm 0.58}$ & 33.28 $_{\pm 0.56}$ & 33.15 $_{\pm 0.36}$ \\
            IFEval & 61.51                                                  & 61.87 & 59.77 $_{\pm 0.63}$ & 59.88 $_{\pm 0.90}$ & 60.11 $_{\pm 0.83}$  & 59.53 $_{\pm 0.83}$  & 59.83 $_{\pm 0.74}$ & 60.43 $_{\pm 1.02}$ & 60.62 $_{\pm 1.42}$ & 60.50 $_{\pm 1.06}$ & 61.01 $_{\pm 0.79}$ & 60.85 $_{\pm 1.07}$ \\
            MATH   & 9.21                                                   & 8.08  & 9.21 $_{\pm 0.72}$  & 9.31 $_{\pm 0.47}$  & 9.35 $_{\pm 0.68}$   & 9.24 $_{\pm 0.52}$   & 9.54 $_{\pm 0.59}$  & 10.02 $_{\pm 0.50}$ & 10.21 $_{\pm 0.53}$ & 9.97 $_{\pm 0.95}$  & 9.70 $_{\pm 0.91}$  & 9.81 $_{\pm 0.40}$  \\
            MMLU-P & 36.67                                                  & 36.22 & 35.91 $_{\pm 0.67}$ & 36.44 $_{\pm 0.27}$ & 36.12 $_{\pm 0.60}$  & 36.21 $_{\pm 0.46}$  & 36.07 $_{\pm 0.29}$ & 35.90 $_{\pm 0.36}$ & 36.21 $_{\pm 0.25}$ & 36.23 $_{\pm 0.19}$ & 36.14 $_{\pm 0.33}$ & 36.03 $_{\pm 0.36}$ \\
            MuSR   & 42.83                                                  & 42.71 & 41.87 $_{\pm 0.78}$ & 42.56 $_{\pm 0.67}$ & 41.90 $_{\pm 0.56}$  & 42.23 $_{\pm 0.83}$  & 42.54 $_{\pm 0.44}$ & 42.65 $_{\pm 1.01}$ & 42.74 $_{\pm 0.55}$ & 42.68 $_{\pm 0.51}$ & 42.78 $_{\pm 0.97}$ & 43.16 $_{\pm 0.64}$ \\
            \bottomrule
        \end{tabular}
    } 
    \caption{Performance of various models across different stress levels for various tasks. Values are averaged over multiple prompts and expressed with their respective standard deviations. For more results, please refer to Table A1 in the Appendix.}
    \label{tab:performance}
\end{table*}

We evaluated the performance of several instruction-tuned LLMs under varying stress conditions using the \textit{StressPrompts} dataset. The models tested included Llama-3-8B-Instruct, Llama-3.1-8B-Instruct, Llama-3-70B-Instruct \cite{llama3modelcard}, Phi-3-mini-4k-Instruct \cite{abdin2024phi3technicalreporthighly}, Qwen2-72B-Instruct, Qwen2-7B-Instruct \cite{qwen2}, and Mistral-7B-Instruct-v0.3 \cite{jiang2023mistral7b}. The generation temperature was set to 0, and specific dialogue tokens were used to ensure consistency.

We utilized a range of benchmarks that assessed emotional intelligence, bias detection, instruction following, reasoning, and mathematical problem-solving. The datasets employed in these evaluations included \textbf{IFEval}~\cite{zhou2023instruction}, \textbf{BBH}~\cite{suzgun2022challenging}, \textbf{MATH}~\cite{hendrycks2021measuring}, \textbf{GPQA}~\cite{rein2023gpqa}, \textbf{MuSR}~\cite{sprague2023musr}, \textbf{MMLU-P}~\cite{wang2024mmlu}, \textbf{EQ-Bench}~\cite{paech2023eq}, \textbf{MMLU}~\cite{hendryckstest2021}, \textbf{TruthfulQA}~\cite{lin2021truthfulqa}, and \textbf{ToxiGen}~\cite{hartvigsen2022toxigen}. The evaluations were conducted using the \texttt{lm\_eval}~\cite{eval-harness} framework with default settings. Baseline prompts used for comparison were \textit{you are a helpful assistant} and \textit{let's think step by step}.

All evaluations were performed on NVIDIA A100 GPUs.  A more detailed description of the experimental setup is provided in the Appendix. 


\vspace{-2mm} 
\subsection{Analysis Under Varying Stress Levels}

The experimental results summarized in Table~\ref{tab:performance} illustrate the effects of varying stress levels induced by \textit{StressPrompts} on the performance of different language models across multiple tasks. Our analysis focuses on the impact of stress on several dimensions, including task performance, model sensitivity, and general trends observed.

In most tasks, moderate stress levels enhance performance, while high stress levels lead to declines, consistent with the Yerkes-Dodson law. This suggests that moderate stress stimulates cognitive engagement, whereas excessive stress overwhelms the system and impairs function.

Complex reasoning and problem-solving tasks, such as \textbf{MuSR} and \textbf{MATH}, exhibit significant performance variations under different stress levels. These tasks benefit from moderate stress but experience marked declines under high stress. For example, Llama-3-8B-Instruct's performance on MATH improves from 0.04 at stress level 1 to 2.93 at stress level 6, demonstrating the positive impact of moderate stress on problem-solving abilities. Similarly, multitask understanding tasks follow this trend, with moderate stress levels enhancing performance. The impact of stress is particularly pronounced in professional-level tasks like \textbf{MMLU-PRO}, where tasks with higher cognitive loads show greater benefits from moderate stress. These findings underscore the unique advantage of \textit{StressPrompt} in addressing complex reasoning and problem-solving challenges. By fine-tuning stress levels, \textit{StressPrompt} can effectively enhance LLMs' performance in tasks requiring high cognitive load, aligning LLM performance with human-like responses under stress.

Different large models exhibit varying sensitivity to stress, with a similar trend observed across multiple models. For instance, Llama-3-8B-Instruct shows substantial improvement in several tasks under moderate stress, while models like Mistral-7B-Instruct-v0.3 display more gradual performance changes. This indicates that model architecture and training specifics play a crucial role in how stress affects performance. While some models, such as Qwen2-7B-Instruct and Phi-3-mini-4k-Instruct, exhibit relatively smaller fluctuations in performance under different stress levels, they are still influenced by stress. These differences may be attributed to varying strategies and preferences during fine-tuning. Overall, while the impact of stress on model performance is evident, the extent and nature of these changes vary depending on the model's training approach.

Figure~\ref{fig:leaderboard_bbh} illustrates the normalized accuracy of various LLMs on subtasks within the \textbf{BBH} benchmark across different stress levels. This benchmark evaluates the cognitive and reasoning abilities of LLMs through tasks such as boolean expressions, causal judgment, date understanding, formal fallacies, geometric shapes and object counting, logical reasoning, and navigation. Our analysis reveals that task complexity significantly impacts the stress level at which peak performance is achieved. Notably, more complex tasks, like logical reasoning with a greater number of objects, tend to reach optimal performance at lower stress levels. For instance, tasks such as \textit{logical\_deduction\_seven\_objects} perform best under less stress compared to simpler tasks like \textit{date\_understanding}. This pattern suggests that higher task complexity imposes a greater cognitive load, making lower stress levels more favorable for maintaining high performance and preventing cognitive overload.

Furthermore, our findings highlight that more powerful models achieve peak performance at lower stress levels, likely due to their advanced capabilities and fine-tuned parameters, enabling them to handle cognitive loads more efficiently under reduced stress. Consistent with the Yerkes-Dodson law, this suggests that LLMs exhibit stress response patterns similar to those of humans, where complex tasks benefit from lower arousal levels to enhance concentration, while tasks requiring endurance may benefit from higher arousal levels to boost motivation. Therefore, the optimal stress levels for LLM performance depend on the nature and complexity of the task, underscoring the importance of adjusting stress levels to match specific task demands.

These observations primarily focus on general cognitive abilities. In subsequent analyses, we will conduct a more detailed examination of emotional intelligence, bias detection, and hallucination. This initial analysis provides a foundational understanding of how stress impacts general task performance, setting the stage for deeper insights into specific cognitive and social competencies.

\subsection{Impact of Stress on Emotional Intelligence, Bias, and Hallucination}

\begin{figure}[h]
    \centering
    \includegraphics[width=.95\linewidth]{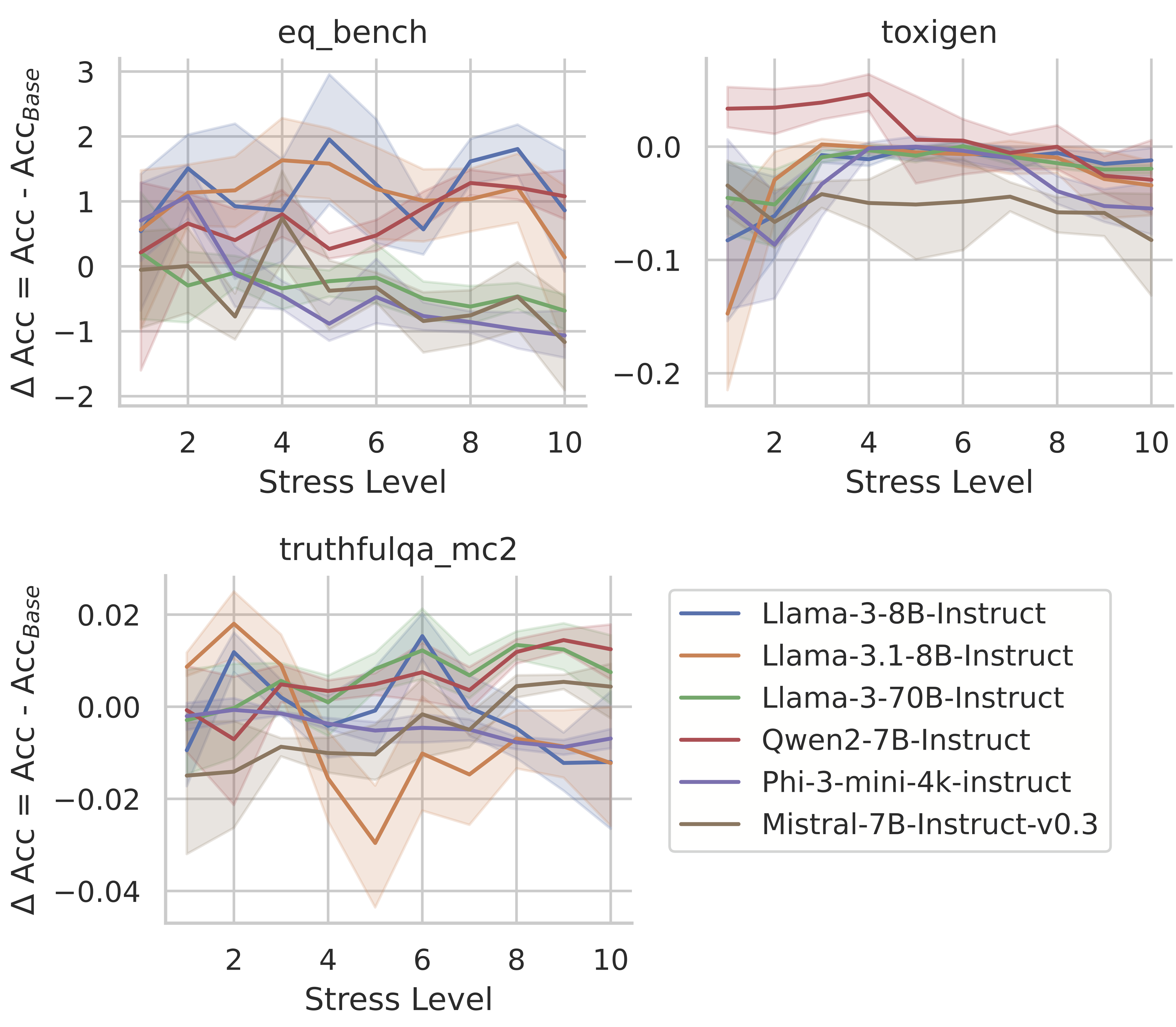}
    \vspace{-3mm} 
    \caption{Performance changes compared to baseline across different stress levels for EQ-Bench, ToxiGen, and TruthfulQA.}
    \label{fig:eq_toxigen_truthfulqa}
    \vspace{-6mm}
\end{figure}

As depicted in Figure~\ref{fig:eq_toxigen_truthfulqa}, the effects of varying stress levels on LLM performance across three datasets—EQ-Bench for emotional intelligence, ToxiGen for bias detection, and TruthfulQA for susceptibility to hallucination—reveal nuanced patterns. For emotional intelligence, models exhibit improved performance under moderate stress, with declines at both low and high stress extremes. This suggests that a balanced level of arousal enhances cognitive engagement without overwhelming the model. 

In contrast, increased stress levels correlate with declining performance in bias detection, indicating that higher stress exacerbates biases. This finding is critical for applications requiring unbiased decision-making, such as content moderation. Regarding hallucination susceptibility, stress has minimal impact, with performance remaining stable across stress levels. This suggests that hallucinations are driven more by intrinsic model factors rather than by stress-induced arousal.

These findings underscore the importance of tailoring stress levels to optimize LLM performance, particularly in tasks demanding high emotional intelligence and fairness. By understanding how stress affects different cognitive and social competencies, we can better align LLMs with human-like responses, enhancing their utility in diverse applications.

\subsection{Visualization of the Effect of Stress on Neural Activity}

\begin{figure}[h]
    \centering
    \includegraphics[width=\linewidth]{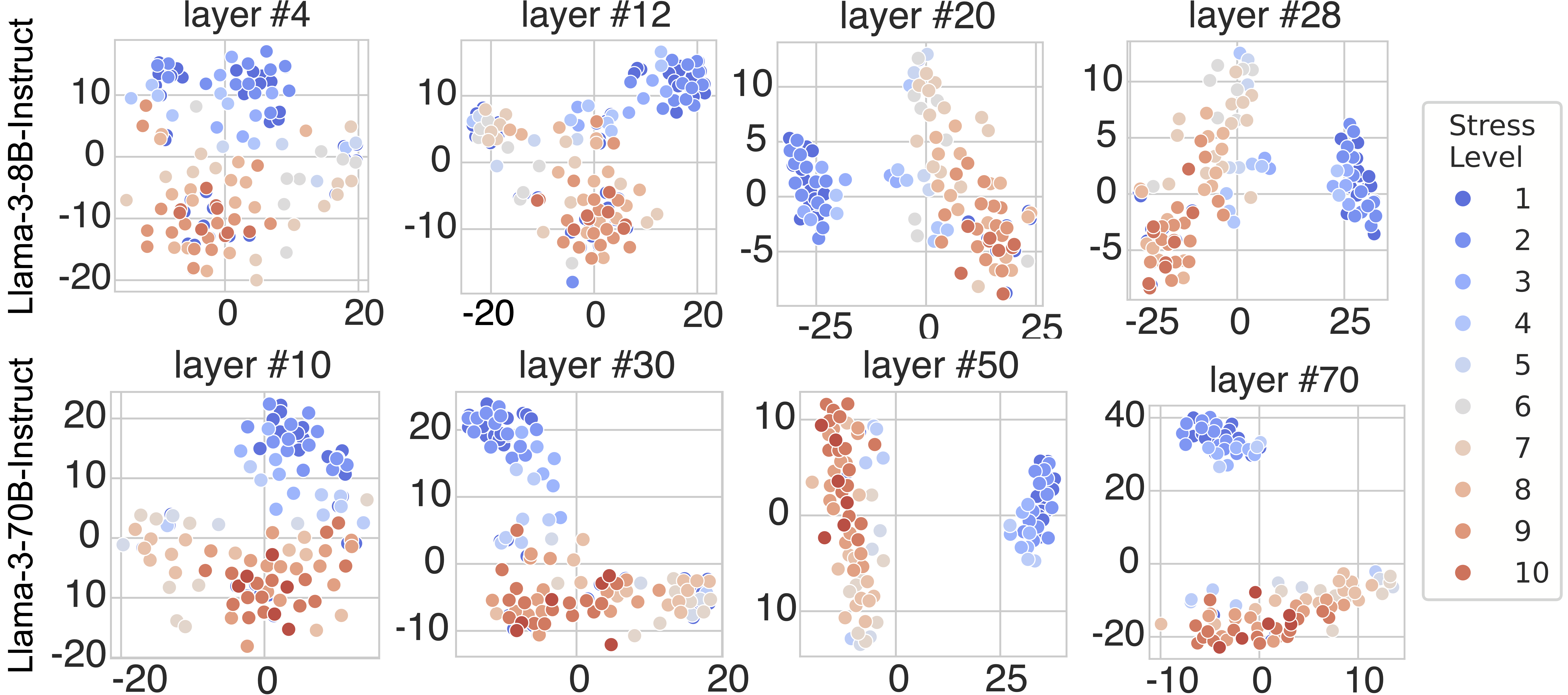}
    \caption{T-SNE visualization of the neural activities of Llama-3-8B-Instruct and Llama-3-70B-Instruct in various layers when processing the last token under different stress levels.}
    \label{fig:tsne}
    \vspace{-3mm}
\end{figure}

\begin{figure}[h]
    \centering
    \includegraphics[width=.9\linewidth]{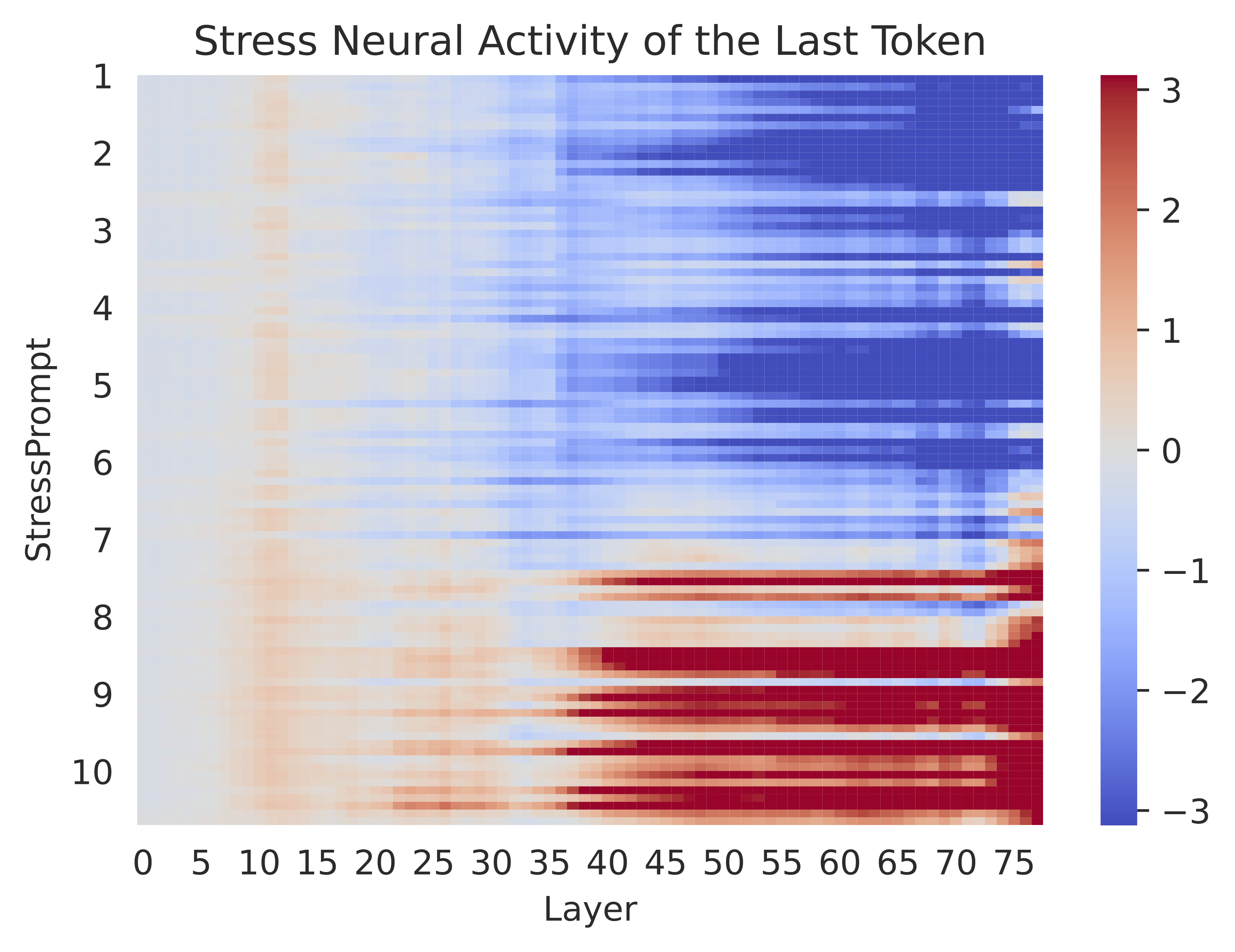}
    \caption{Heatmap of neural activity of the last token across all layers for various stress levels in Llama-3-70B-Instruct.}
    \label{fig:heatmap}
    \vspace{-6mm}
\end{figure}

To gain insights into how LLMs respond to different stress levels, we visualized their neural activity. As shown in Figure~\ref{fig:stress_scanner}, the neural activity of the last token when inputting \textit{StressPrompt} effectively reflects the induced stress. We conducted an experiment using T-SNE to visualize the neural activities of LLMs across various layers, as depicted in Figure~\ref{fig:tsne}. The results indicate that initial layers are unable to distinguish between stress levels, whereas deeper layers can classify prompts into low-stress and high-stress categories, indicating a higher sensitivity to stress in these layers.

Furthermore, we performed a stress scan on the last token of all prompts, illustrated in the heatmap in Figure~\ref{fig:heatmap}. This visualization captures neural activity across all layers for various stress levels, revealing significant changes in deeper layers. Specifically, deeper layers exhibit more pronounced differences between low and high-stress levels, underscoring their critical role in detecting and responding to stress. Research indicates that higher cognitive regions of the human brain, such as the prefrontal cortex, show significant activity changes under stress, particularly during complex and high-pressure tasks. Our findings suggest that the deeper layers of LLMs exhibit similar sensitivity to stress, reflecting the analogous impact of stress on both human brains and LLMs.

\vspace{-3mm}
\section{Conclusion}

Our study demonstrates that LLMs exhibit performance patterns closely resembling those of humans under varying stress levels. By constructing the \textit{StressPrompt} dataset, we found that LLMs can map human-like relationships between stress and task performance. Moderate stress enhances capabilities such as reasoning and instruction following, while excessive stress impairs tasks like bias detection. This mapping enables LLMs to emulate human strategies in problem-solving, adapting stress levels to optimize performance. These findings suggest that large models have captured and operationalized human-like stress-performance dynamics, paving the way for more resilient and adaptive AI systems.

\section{Acknowledgments}

This research was funded by the Central Government-Guided Local Special Fund within the Beijing Science and Technology Program (Grant No.Z241100001324005).
   



\bibliography{aaai25}

\end{document}